\documentclass[preprint,amsmath]{revtex4-1}
\usepackage{graphicx}
\usepackage{caption}
\usepackage{subcaption}
\usepackage{hyperref}

\def\bq{\begin{eqnarray}}
\def\eq{\end{eqnarray}}
\def\nn{\nonumber}

\begin{document}

\title{One-photon pair production on the expanding de Sitter spacetime}

\author{Robert Blaga}
\email{robert.blaga@e-uvt.ro}
\affiliation{West University of Timi\c soara,\\
V.  P\^ arvan Ave.  4, RO-300223 Timi\c soara, Romania
}

\date{\today}

\begin{abstract} We study the one-photon scalar pair production QED process on the expanding de Sitter spacetime. Using perturbation theory, we obtain the transition probability and study its properties as a function of the expansion parameter $\omega$. On flat space the process is forbidden by energy-momentum conservation. It is expected that for a dynamical background there is an energy exchange correlate to the strength of the gravitational field. We use momentum space plots and compute the mean production angle to illustrate this. We show that the mean angle grows with $\omega$, but also find that in the flat limit the fall-off is unexpectedly slow. To investigate this further we obtain the probability around different angular configuration, at leading order in $m/\omega$, and find that the $\omega$ dependence at small angles is very weak. We comment on the possible astrophysical implications.
\end{abstract}

\pacs{04.62.+v}
\maketitle

\section{Introduction}
In the absence of a complete theory of quantum gravity, the theory of quantum fields on curved spacetimes remains our best tool for investigating nature at the most fundamental level.\,One of the most important results that have arisen from this approach is the phenomenon of gravitational particle production. In essence what this entails is that in a dynamic spacetime, pairs of particles can be produced purely out of the vacuum. Starting with the original paper by Schrodinger \cite{Sch}, through the seminal work by Parker \cite{P1,P2,P3}, up till the numerous present day studies, a considerable body of work has accumulated regarding the subject. Furthermore, the problem of self-interacting (scalar) fields on curved backgrounds has been investigated quite extensively. The main interest is in the renormalization on different backgrounds \cite{BF1,TO,BF2} and one of the important questions addressed is wether the different spacetimes are stable under perturbations (see for example \cite{ST,A1} and references therein). An adjacent topic, but which has been only sparsely investigated in the literature, is the effect of the dynamical background on the mutual interaction of quantum fields and inversely the contribution of the mutual interaction to particle production. Considering that in a physical setting the gravitational production is always accompanied by interaction processes, the study of the mutual interaction is imperative for obtaining a complete picture.   \par
Birrell, Davies and Ford \cite{BDF} are the first to analyze the effects of field interactions upon particle creation in FRW universes. They argue that the interaction may significantly affect and, for bounce model universes, even dominate the pure gravitational production. The study of QED processes was undertaken by Lotze in expanding FRW universes \cite{L3,L1,L2}, Tsaregodtsev et.al. in a radiation dominated universe \cite{TS1,TS2} and Cotaescu,Crucean et.al. in de Sitter space \cite{AS4,CB1,BL}. Furthermore there are a number of papers dealing with the quantum field theoretic treatment of radiating charges in different FRW universes \cite{N1,N2,H2}, at leading order in the WKB approximation or the perturbation theory. \par
The studies involving de Sitter space (dS) are particularly relevant due to the cosmological significance of the spacetime. In this paper we present a detailed analysis of the transition probability for one-photon scalar particle pair production on the expanding patch of dS. We work in the in-out formalism, with the S-matrix approach. In order to focus on the contribution of the interaction to the transition probability, we consider the final vacuum state to be identical to the \emph{in} vacuum, and define the notion of \emph{particle} with respect to the in modes. The resulting amplitudes are sometimes called in-in amplitudes in the literature. The notion of particle is defined with respect to the in-modes. The resulting probability is interpreted also in terms of \emph{added-up} probabilities, introduced in ref.\cite{AS1} and described in Appendix \ref{C}. \par
The paper is structured as follows. Section I contains the necessary ingredients for setting up the perturbation theory for the scalar QED on de Sitter space. In section II we obtain the transition probability and perform an extensive analysis as a function of the expansion parameter. In section III we summarize the main results. We work in natural units where $\hbar=c=1$.

\section{Preliminaries}
The expanding Poincare patch of the de Sitter spacetime is described by the following metric, written in both planar and conformal coordinates \cite{LH}:
\begin{eqnarray}
ds^{2}&=&dt^{2}-e^{2\omega t}d\vec{x}\,^{2} \\
&=&(\omega t_{c})^{-2}(dt_{c}^{2}-d\vec{x}^{\,2}), \nn
\end{eqnarray}
where $\omega$ is the Hubble parameter and the conformal time is defined as $\omega t_{c}=-e^{-\omega t}$, $t_{c}\in(-\infty, 0)$. \par
We consider a massive scalar field minimally coupled to gravity. The Klein-Gordon (KG) equation in this geometry is:
\bq
(\partial^2_t - e^{-2\omega t} \Delta + 3\omega\partial_t + m^2)\, \varphi (x,t) = 0
\eq
The positive frequency plane-wave solutions of the KG equation are \cite{AS2,NA}:
\begin{equation}
f_{\vec{p}}\,(x)=\frac{1}{2}\sqrt{\frac{\pi}{\omega}}\frac{e^{-3\omega t/2}}{(2\pi)^{3/2}}e^{i\pi\nu/2}H^{(1)}_{\nu}\left(\frac{p}{\omega}e^{-\omega t}\right)e^{i\vec{p}\,\vec{x}}, \label{a1}
\end{equation}
where $H^{(1)}_{\nu},H^{(2)}_{\nu}$ are the Hankel functions with index $\nu=i\sqrt{\mu^2-\frac{9}{4}}$, and  $m =\omega\mu$ is the mass of the scalar field. These modes define the Bunch-Davies vacuum \cite{BD}.\par
The Maxwell field is quantified in the Coulomb gauge \cite{AS3}, given by $A_{0}=0$, $(\sqrt{-g}A^{i})_{;i}=0$. Because the field is conformal, the mode functions can be straightforwardly obtained from the flat space solutions \cite{AS3}:
\begin{equation}
w^{i}_{\vec{k},\lambda}(x)=\frac{1}{(2\pi)^{3/2}}\frac{1}{\sqrt{2k}}\,e^{-ikt_{c}+i\vec{k}\,\vec{x}}\,\vec{\varepsilon}_{\lambda}(\vec{k}\,)\,e^{-2\omega t}, \label{a2}
\end{equation}
where $\vec{\varepsilon}_{\lambda}(\vec{k}\,)$ are the polarization vectors. \par
With the above expressions (\ref{a1}) and (\ref{a2}), the usual mode expansion for the free scalar and Maxwell fields can be written as:
\bq
\varphi(x) &=& \int d^3p\, \left( a(\vec{p}\,) f_p(x) +\, b^\dag (\vec{p}\,) f^*_p(x)    \right) \\
A^i(x) &=& \sum\limits_{\lambda}\ \int d^3k\, \left( c_\lambda(\vec{k}\,)\, w^{i}_{\vec{k},\lambda}(x) + c^*_\lambda(\vec{k}\,)\, w^{i\,*}_{\vec{k},\lambda}(x) \right) \\
A^0(x) &=& 0, \qquad \qquad  i=1,2,3, \nn
\eq
where $a^\dag, b^\dag$ and $c^\dag$ are creation operators for scalar particles, antiparticles and photons, respectively. \par
 We use a general prescription for interacting field theories \cite{BD} and consider QED processes in the first order of perturbations.
  The 1st order term in the perturbation expansion of the $S$ matrix is given by:
\begin{eqnarray}
S^{(1)}= -e\int d^{4}x \sqrt{-g}(\varphi^{+}(x)\stackrel{\leftrightarrow}{\partial}_{\mu}\varphi(x))A^{\mu}(x),
\end{eqnarray}
where $\sqrt{-g}=e^{3\omega t}$ and $f\stackrel{\leftrightarrow}{\partial}g = f(\partial g)-g(\partial f)$ is the bilateral derivative. We have omitted the four-point interaction term, which is second order in the coupling constant, because it does not contribute to the processes studied here.\\
The transition amplitude for these first order processes is defined as:
\begin{eqnarray}
\mathcal{A}\,_{i\rightarrow f}=  \langle out\,|S^{(1)}|\,in\rangle.
\end{eqnarray}
With this, we have all the preliminaries in place in order to study tree-level processes. In our previous work \cite{BL} we have considered the emission of photons by a scalar particle, and here we investigate the process of one-photon pair production.

\section{One-photon pair production}
\begin{figure}
\includegraphics[width=3.5in]{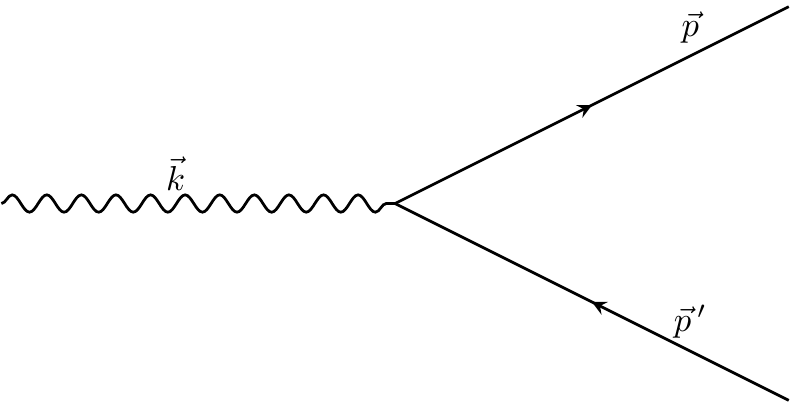}
\caption{Feynman diagram - pair production}
\label{1}
\end{figure}

Within the framework introduced in the previous section, we consider the process of scalar particle pair creation by a single photon, illustrated by the Feynman diagram $(\ref{1})$. \par
In flat spacetime this process is forbidden at tree-level. This can most easily be seen by changing to the center of mass frame of the resulting pair of particles. Due to energy-momentum conservation, the momentum of the photon would then have to be vanishing while at the same compensating for the rest energy of the massive particles, at the least. An analogous situation is instead the second order process of one-photon pair creation in an external field, commonly in the field of a nucleus \cite{Hu} or in a strong magnetic field \cite{DH}. The first is relevant because of the possibility of experimental confirmation, while the second is relevant in astrophysical settings. In our case the background plays the role of the external (gravitational) field. \par

This process is also of interest from another point of view. The interpretation of transition amplitudes and more generally of what represents a measurable quantity on a non-asymptotically flat spacetime is still a matter of some debate. It is thus desirable to deal with quantities whose interpretation is insensitive to the definition of the out state for example. The concept of added-up probability, introduced by Audretsch and Spangehl \cite{AS1}, is such a quantity. \par
In this approach one calculates the probability for transition between two states of the Maxwell field with different occupation numbers, irrespective of what happens to the massive field. This is accomplished by summing over the complete Fock space of the scalar field in the final state. This method can be employed because the Maxwell field is conformal and thus the vacuum state is the same for all times, i.e. photons can not be created freely from the background. A brief summary of the method of added-up probabilities can be found in Appendix \ref{C}.\par
For the specific case of pair production by a photon, the added-up probability means the transition from the in-state containing a photon (an no scalar particles) to a state with no photons (vacuum of the Maxwell field), summed over all possible outcomes for the scalar field. In other words, the probability that a photon "disappears", regardless of "into what".\ What is remarkable in this case is that the diagram (\ref{1}) is the only one that contributes to the added-up probability \cite{L1,L2}. This is not true for the time-reversed process (see Appendix \ref{C}) . The added-up probability can thus be used as a Rosetta stone to translate between in-in and in-out probabilities. \par

\subsection{Transition probability: expression}
The transition amplitude which describes the one-photon scalar pair production has the following expression:
\bq
\mathcal{A}(\vec{p},\vec{p}\,',\vec{k}) &=& \langle\, 1_{\,(\vec{p}\,)}\, ,\tilde{1}_{\,(\vec{p}\,')}\, \vert\ S^{\,(1)}\ \vert \,1_{\,(\vec{k},\lambda)} \, \rangle\ \nonumber\\
&=&-e\int d^{\,4}x\,\sqrt{-g}   \left(f^*_{\vec{p}}\,(x)\stackrel{\leftrightarrow}{\partial}_{i}f^*_{\vec{p}\,'}(x)\right)\,w^{\,i}_{\vec{k},\,\lambda}(x).\label{p1}
\eq
Introducing the solutions $(\ref{a1})$ for the scalar field  and $(\ref{a2})$ for the Maxwell field, the transition amplitude becomes:
\bq
\mathcal{A}(\vec{p},\vec{p}\,',\vec{k})
&=& \,\delta^{3}( \vec{p} +\, \vec{p}\,' - \vec{k}\,)\ \frac{ie\pi(\vec{p}\,'-\,\vec{p}\,)\cdot \vec{\varepsilon}\,(\vec{k}\,)}{(2\pi)^{3/2}\sqrt{32k}}\ e^{-i\pi \nu}\,I^{(2,2)}_{\nu}(p,p',k),
\label{p2}
\eq
where
 \begin{eqnarray}
 I_{\,\nu}^{(a,b)}(p,p',k) = \int_0^\infty d\eta\ \eta\ H^{(a)}_\nu(p'\eta)\,H^{(b)}_\nu(p\,\eta)\ e^{ik\eta}.\label{ti}
 \end{eqnarray}
The form (\ref{ti}) for the temporal integral was arrived at by changing the integration variable to $\eta = - t_c$. \par
The tree-level (scalar) QED amplitudes on dS space all have the same structure, with a momentum-conserving delta function arising from the spatial integral as a consequence of spatial translation invariance, and with the temporal integral contributing the non-trivial physics. \par
The calculation of the integral $I^{(2,2)}_{\nu}$ is given in Appendix $(\ref{A.1})$. From the form (\ref{E9}) of the amplitude we observe that it depends only on the ratio $\mu = m/\omega$ and the involved momenta. Unless stated otherwise, in the following we shall consider unit mass in the formulas and plots. \par
We add to the integral (\ref{ti}) an exponential factor $e^{-\epsilon\eta}$ that acts as a switch-off for the interaction for large times, the decoupling time being of order $1/\epsilon$. The \,$\epsilon$ is a small positive parameter that reinforces the convergence of the integral. Effectively we consider it to be a small but finite constant in our calculations, and discuss the vanishing limit at the end of the paper. \par

The partial probability, averaged over the photon polarizations, is obtained as: \par
\bq
\mathcal{P}(\vec{p},\vec{p}\,',\vec{k}) &=& \frac{1}{2}\,\sum_{\lambda = \pm 1}\  \left\vert\, \mathcal{A}(\vec{p},\vec{p}\,',\vec{k})\ \right\vert^{\,2}
\label{PP1}
\eq
The delta term can be handled, following the prescription from flat space scattering theory, by writing
\bq
\left\vert (2\pi)^3\,\delta^{3}\left(\sum \vec{p}\,\right)\right\vert^{\,2} =(2\pi)^6\, \delta^{(3)}(0)\,\delta^{3}\left(\sum \vec{p}\,\right) =(2\pi)^3\ V\ \delta^{3}\left(\sum \vec{p}\,\right),
\eq
where $V$ is the comoving volume. The physical volume is $V_{phys} = Ve^{3\omega t} = (\omega\eta)^3\,V$. One then usually considers the probability per unit volume. In the following we will shorthand this quantity with just the probability, remembering that what we really mean is in fact the probability per unit comoving volume.\par
The polarization term can be easily obtained by making use of the relations $ \vec{p}\,' = \vec{k} - \vec{p}$ (momentum conservation) and $\vec{k}\cdot \vec{\varepsilon}_\lambda(\vec{k}) = 0$.
\bq
  \sum_\lambda \left\vert(\vec{p}\,' - \vec{p}\,)\cdot \vec{\varepsilon}_\lambda(\vec{k}\,)\right\vert^{\,2}  &=&\ \left\vert 2\,\vec{p} \cdot \vec{\varepsilon}_\lambda(\vec{k}\,) \right\vert^{\,2}
  \label{pol} \\
 &=& 4\left(\vec{p}\,^2 - \frac{(\vec{p}\cdot\vec{k}\,)^2}{k^2}\right) \nn\\
 &=& 4p^{\,2} \sin^2\theta \nn\\
 &=& \frac{4p^{\,2}p'^2 \sin^2\chi}{k^2}
 \nn
\eq
where in (\ref{pol}) we recognize the projection of the momentum onto the plane defined by the polarization vectors (and perpendicular to the direction of $\vec{k}\,$), and $\theta$ represents the angle between $\vec{p},\vec{k}$, while $\chi$ represents the angle between $\vec{p},\vec{p}\,'$. \\
Gathering all terms and using the notations from Appendix $(\ref{A.1})$, the probability becomes:
\bq
\label{PP}
\mathcal{P}(\vec{p},\vec{p}\,',\vec{k}) &=& \delta^{3}(\vec{p}+\vec{p}\,' - \vec{k}\,)\ \ \frac{e^2\pi^2}{16}\,\frac{{p'}^2 p^2 \sin^2\chi}{(2\pi)^6 \,k^3}\ \label{P1}\\
&\times& \left\vert g_{\,\nu}(p,p',k) + g_{-\nu}(p,p',k)+ h_{\nu}(p,p',k)+ h_{-\nu}(p,p',k)\right\vert^{\,2}.\nonumber
\eq
The quantity (\ref{PP}) is to be interpreted as the momentum space distribution of the total probability:
\bq
\mathcal{P}(\vec{p},\vec{p}\,',\vec{k}) = \frac{dP}{d^3p\,d^3p\,'},
\eq
from where the total probability is defined as:
\bq
P(k) = \int d^3p\,d^3p\,'\ \mathcal{P}(\vec{p},\vec{p}\,',\vec{k})
\eq
We shall also consider the following the quantity
\begin{equation}
\tilde{\mathcal{P}}(k,p,\theta) = \int d^3p\,'\ \mathcal{P}(\vec{p},\vec{p}\,',\vec{k}). \label{P2}
\end{equation}
Notice that the delta function makes the above integral trivial. \par
 For later use, we obtain the expression of the probability for the particular value $\mu = \sqrt{2}$, where the Hankel functions have simple analytical forms (\ref{H12}) and the temporal integral can be readily evaluated:

\bq
\label{IC}
I^{(2,2)}_{\pm\frac{1}{2}}(p,p',k) &=& \int d\eta\,\eta\,\sqrt{\frac{2}{\pi p'\eta}}\,e^{-ip'\eta}\,\sqrt{\frac{2}{\pi p\,\eta}}\,e^{-ip\,\eta}\ e^{ik\eta} \\
&=& \frac{2}{\pi} \frac{1}{\sqrt{pp\,'}} \int d\eta\,\eta\ e^{-i(p\,'+p-k)\eta} \nn \\
&=& \frac{2}{\pi} \frac{1}{\sqrt{pp\,'}} \frac{i}{p\,'+p-k} \nn
\eq
The probability thus becomes:
\bq
\tilde{\mathcal{P}}(k,p,\theta) = \frac{1}{(2\pi)^6}\frac{e^2}{4k}\,\frac{p}{p'}\,\frac{\sin^2\theta}{(p+p'-k)^2+\epsilon^2}\ ,
\eq
where we have restored the $\epsilon$ parameter for clarity, and it is understood that $p' = \sqrt{p^2 + k^2 - 2pk\cos\theta}$. The simple form of the probability in this case makes it particularly suitable for numerical and analytical manipulations, and will be exploited accordingly in the following. \par
We note that for this particular case the mode functions reduce to the flat-space plane-waves, and the scalar field effectively behaves as a massless conformally coupled field. One would then expect that the probability be null as in the flat space case. The probability is indeed equal to the flat space one, but neither of them are in fact null. This can be seen by identifying the conformal time $\eta$ with the Minkowski time in eq. \ref{IC}. We observe that the integration is only over the semi-infinite interval. This is because the expanding patch of dS is only \emph{locally} conformal to Minkowski space. The integration up to a finite time in flat space can be understood as a sudden decoupling of the fields at a finite time. This would then lead to transient effects giving the probability a non-null value. Translating back to the expanding dS space, what in flat space was due to (unphysical) transient effects, now represents the physical probability rendered so by the nature of the spacetime.
\subsection{Transition probability: analysis}
The probability can be evaluated analytically for $m \gg \omega$ (weak gravitational regime), with the aid of an approximation. From the expression (\ref{P1}), the $h_{\pm\nu}$ terms which contain an Appell $F_4$ function are the ones that resist analytical manipulation. We have found that the approximation
\bq
\label{F4}
F_4(1,3/2,1+\nu,1-\nu,x,y) \simeq 1,
\eq
works very well for large $\mu$. A more detailed discussion can be found in Appendix (\ref{A.2}). \par
An alternative option is to numerically integrate the temporal integral in (\ref{p2}), which we shall do for the domain $m \sim \omega$ (strong gravitational regime), where the above approximation is not accurate. Finally, we will make use of the analytically evaluated probability (\ref{P2}) for the case $\mu = \sqrt{2}$. \par
\begin{figure}[h]
\centering
\begin{subfigure}[b]{0.5\linewidth}
\centering
\includegraphics[width=3in]{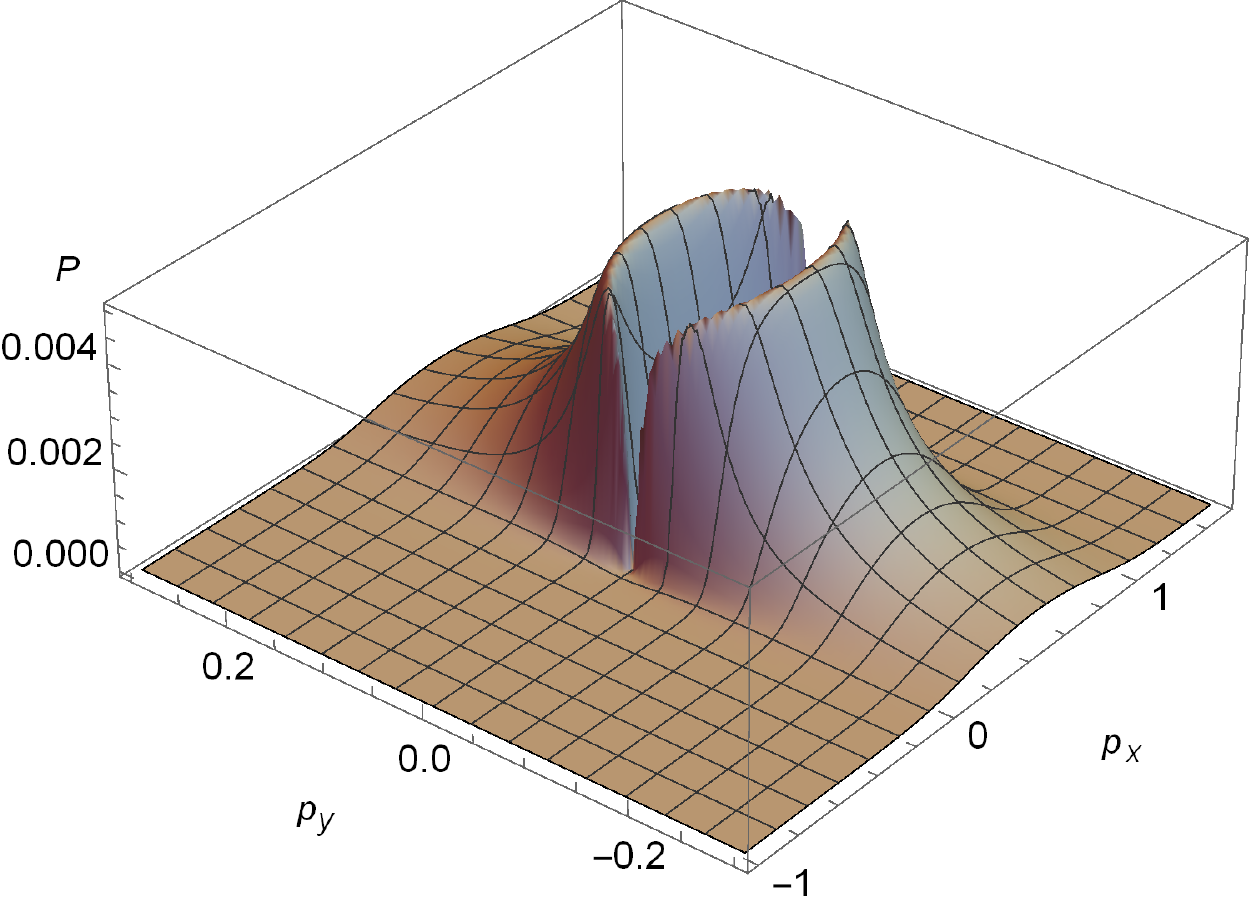}
\caption{$\mu = \sqrt{2}$}
\label{2a}
\end{subfigure}%
\begin{subfigure}[b]{0.5\linewidth}
\centering
\includegraphics[width=3in]{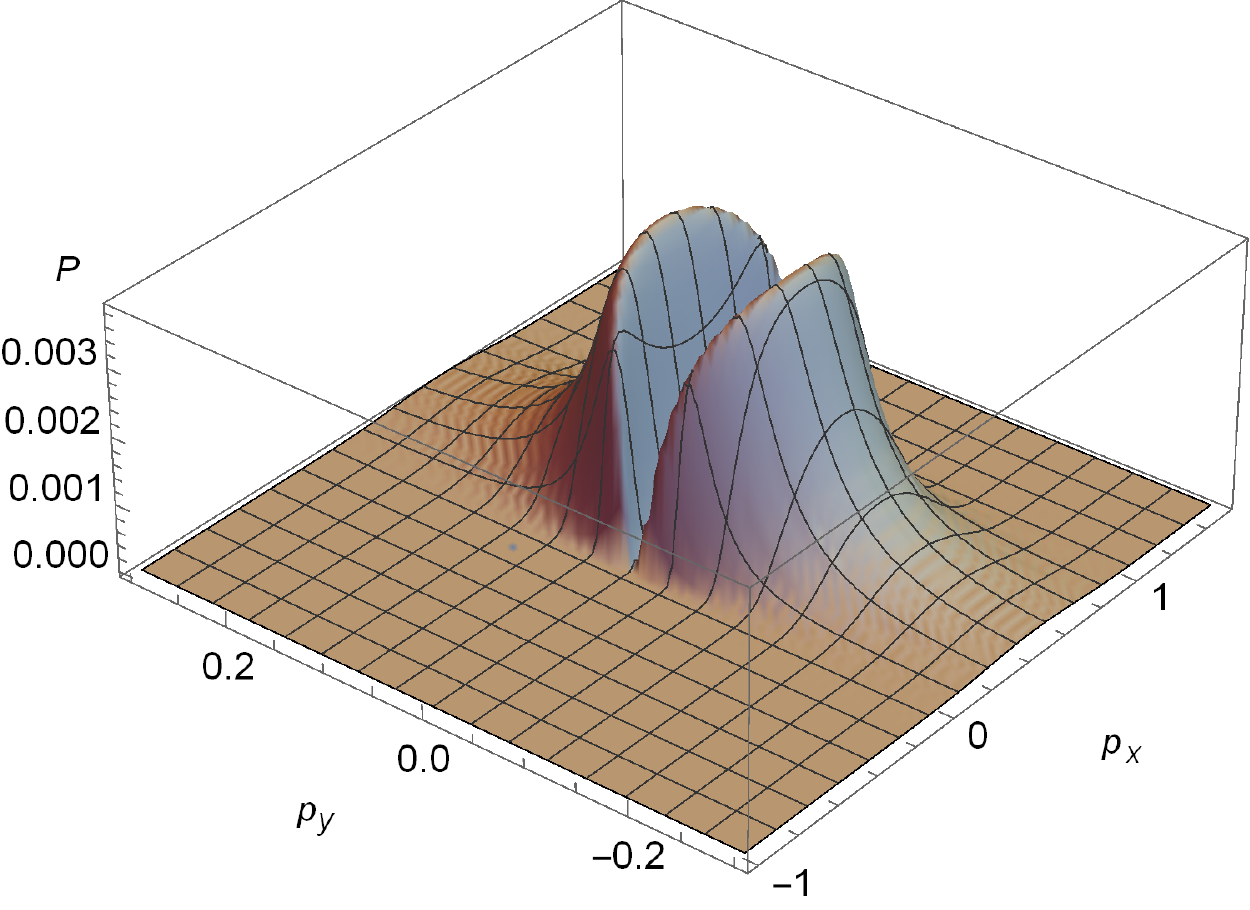}
\caption{$\mu = 2$}
\label{2b}
\end{subfigure}
\begin{subfigure}[b]{0.5\linewidth}
\centering
\includegraphics[width=3in]{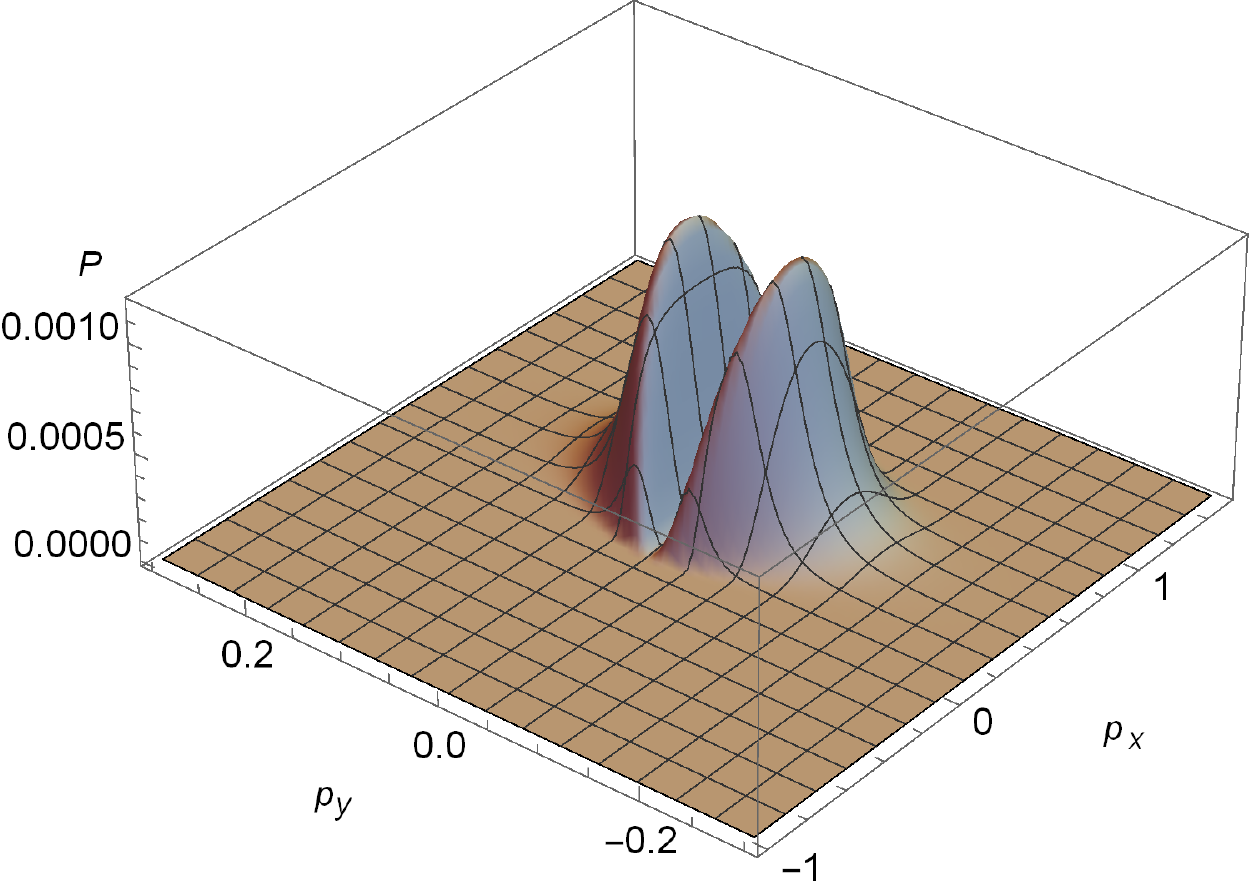}
\caption{$\mu = 5$}
\label{2c}
\end{subfigure}%
\begin{subfigure}[b]{0.5\linewidth}
\centering
\includegraphics[width=3in]{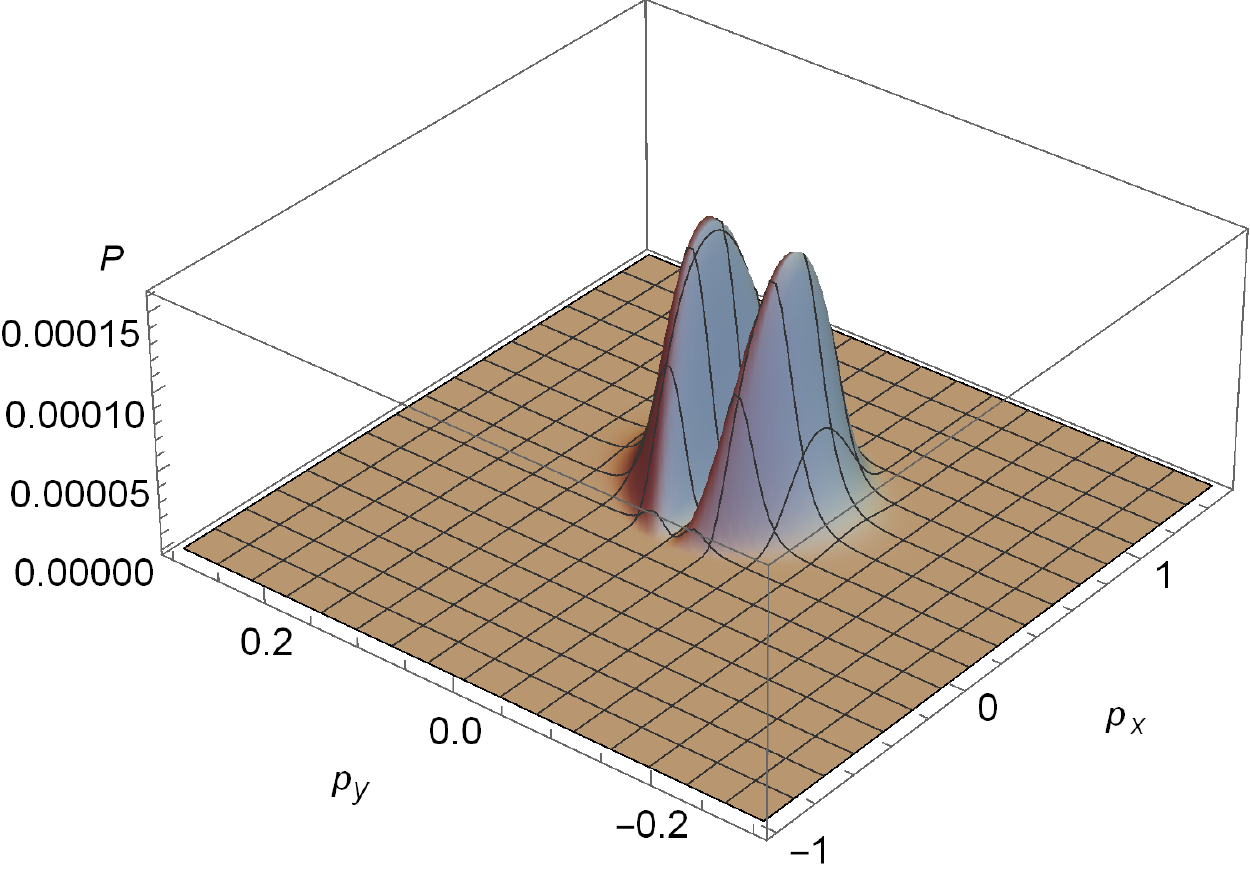}
\caption{$\mu = 10$}
\label{2d}
\end{subfigure}
\caption{\small Momentum space surface plots of the transition probability, in cartesian coordinates, for k=1 and $\epsilon = 10^{-2}$. The x-axis is taken in the direction of the photon momentum $\vec{k}$. The probability being invariant under rotations around the direction of $\vec{k}$, the problem reduces to scattering in a plane.}
 \label{2}
\end{figure}
\par
The probability is best visualized in cartesian momentum space surface plots, with $p_x$ denoting the direction parallel to the photon momentum and $p_y$ the perpendicular. As can be seen from fig.(\ref{2}), the probability takes significant values only in the interval $p_x \in (0,k)$ and at small angles (small $p_y$ component), falling abruptly for negative and large momenta. The overall conclusion to be drawn is that even tough there is the possibility of energy exchange with the dynamic background, the probability is peaked in the vicinity of the energy conserving case ($\theta\simeq \chi \rightarrow 0, k \simeq p+p'$). To get an intuitive picture one can think of the configuration of momenta in Compton scattering in flat space. In that case the scattering at backward angles is forbidden by the simultaneous energy-momentum conservation law, as is the case where the photon gains momentum as a result of the process. In the somewhat analogous situation for our process of pair production in de Sitter space, the backward "scattering" and production of large momentum pairs, although not forbidden, are heavily suppressed. On the other hand, a significant relative increase in the probability of production at intermediate angles, corresponding to production in the forward quadrants, is clearly visible as the strength of the gravitational field increases. Also, notice that while in a weak field the probability has a maximum at $p\simeq p'\simeq k/2$, in the strong field case the probability is the same throughout the interval $(0,k)$. \par
\subsection{Mean production angle}
\label{sec 3.3}
To characterize the angular behaviour of the probability with varying strength of the gravitational field, we calculate the mean production angle for various values of the expansion parameter. If we consider the transition probability as the distribution function for the variable $\theta$, we can obtain the mean production angle as:
\begin{equation}
\langle \theta \rangle = \frac{ \iint \theta\,\tilde{\mathcal{P}}(k,p,\theta)\,p^2dp\, \sin\theta d\theta  } { \iint \tilde{\mathcal{P}}(k,p,\theta)\,p^2dp\,\sin\theta d\theta  }
\end{equation}
The straightforward calculation of the mean angle is not possible however because the probability has an UV divergence. This is highly unexpected, especially considering that in the flat limit the mode functions reduce to the familiar plane-waves, as was shown in ref.\cite{CP}, and the probability vanishes accordingly. To trace the origin of this divergence we note that in the ultra-relativistic case $(p \gg m, p\eta \gg 1)$, the mode functions reduce to the flat space plane waves, but with time coordinate $\eta$.
\begin{figure}[h]
\centering
\begin{subfigure}[b]{0.5\linewidth}
\centering
\includegraphics[width=3in]{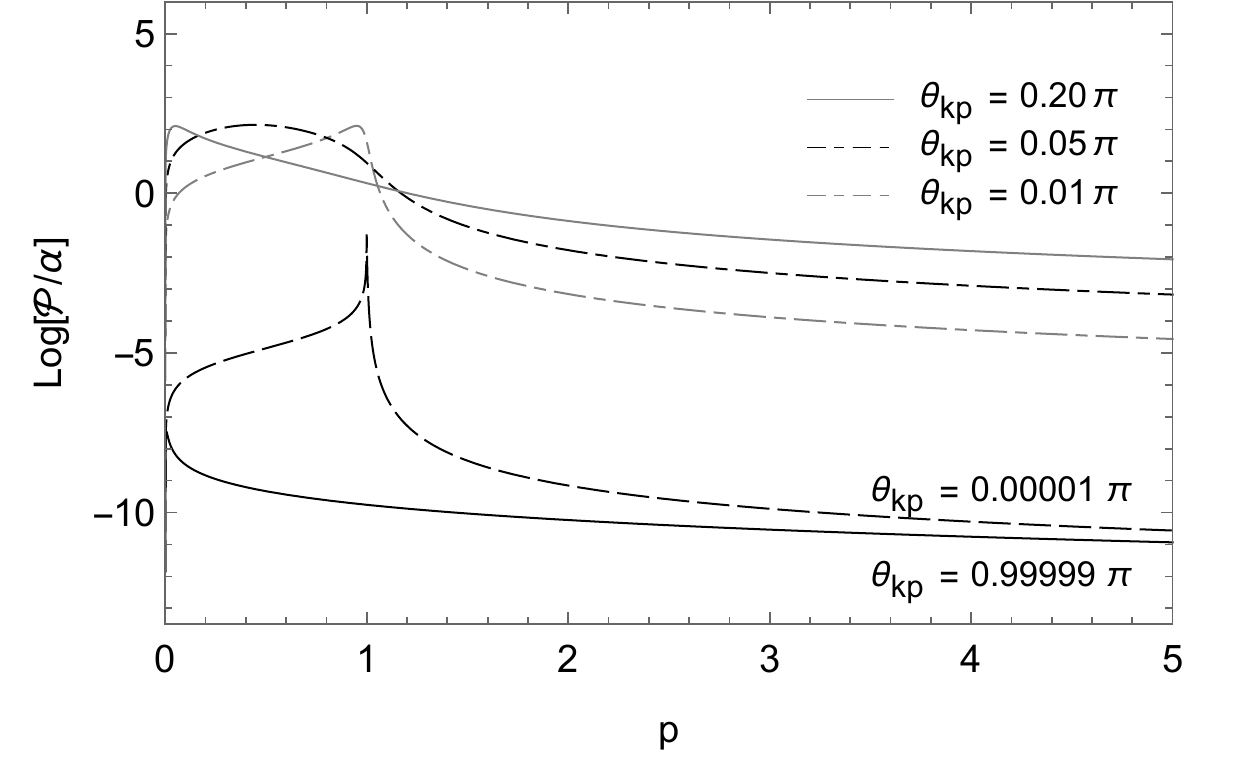}
\caption{}
\label{3a}
\end{subfigure}%
\begin{subfigure}[b]{0.5\linewidth}
\centering
\includegraphics[width=3in]{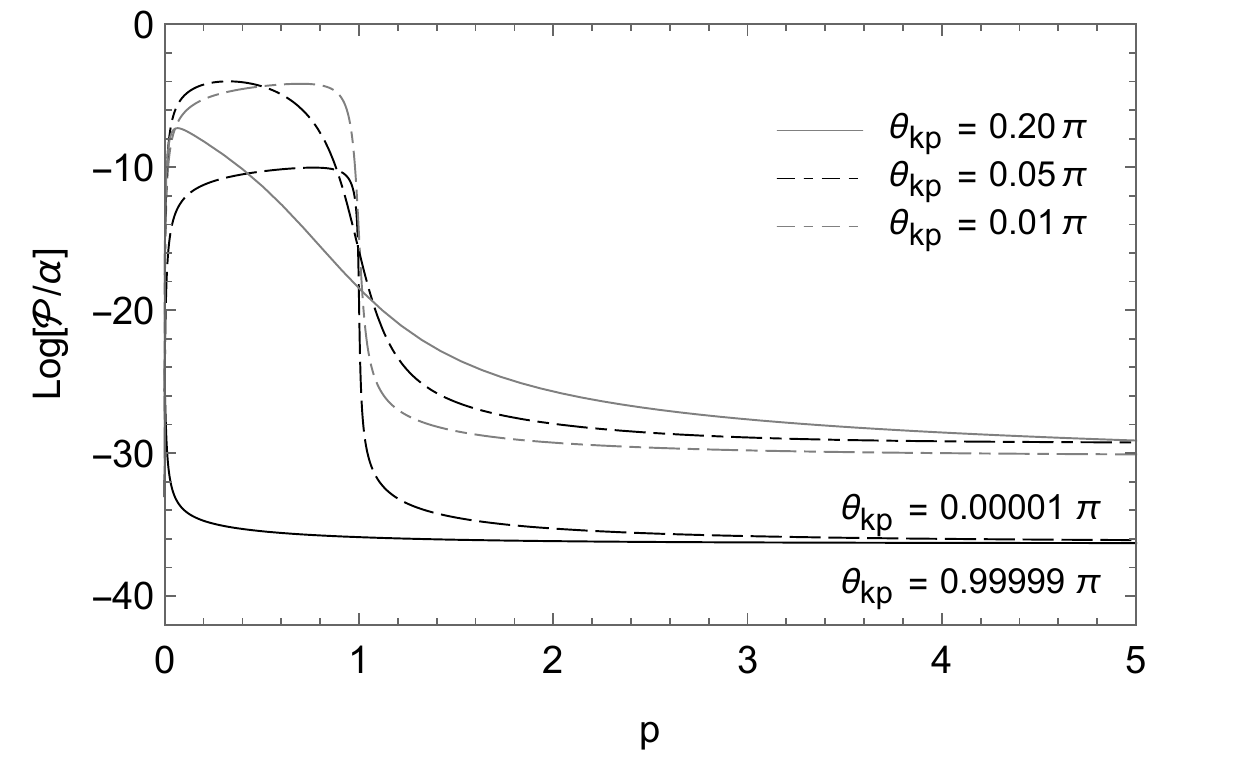}
\caption{}
\label{3b}
\end{subfigure}
\caption{\small The momentum distribution of the probability for $\mu=\sqrt{2}$ (a) and $\mu=10$ (b), for $k = 1$  and various angles. Note the sudden drop of the probability for $p > k$, and also the many orders of magnitude difference in the large $p$ behaviour for the two cases.}
 \label{3}
\end{figure}
One would expect that the amplitudes then reproduce the flat space ultra-relativistic amplitudes for the analogous process in an external field, for example. As we have argued above however, because the range of the conformal time is restricted to the \emph{semi}-infinite axis, the amplitude will actually be similar to that in flat space, but where the interaction has been suddenly decoupled at time $t=0$. This would then lead to transient effects and an UV divergence when integrated over the momenta of the particles, as was reported also in ref.\cite{Nk}. \par
In order to extract a quantity which gives at least qualitative information about the angular behaviour, we effectively integrate the momentum up to a maximum value $p_{max}$. Observing that the probability is highest in the interval $(0,k)$, as can be seen from fig.(\ref{2}) and (\ref{3}), we consider it sufficient to take $p_{max} \sim k$ in order to illustrate the variation of the mean production angle. Increasing $p_{max}$ accentuates further the increase of the mean angle. Listed below is a table with the results obtained for the mean angle, for different values of $p_{max}$ and $\epsilon$: \par
\begin{table}[h]
\centering
\begin{tabular}{|c|r|r|r|r|}
  \hline
  \, & $\epsilon=0.01$ & $\epsilon=0.01$ & $\epsilon=0.001$ & $\epsilon=0.001$ \\
  \,$\mu$ & $p_{max}=2k$ & $p_{max}=10k$ & $p_{max}=2k$ & $p_{max}=10k$ \\ \hline
   $10$ &  6.76 &6.76 & 3.53 & 3.53 \\
   $5$ & 9.40 &  9.54& 5.05 & 5.37 \\
   $2$ & 20.40 & 23.90 & 25.40 & 31.25\, \\
   $\sqrt{2}$ & 45.11 & 48.45 & 35.66 & 60.40 \\
   \hline
\end{tabular}
\caption{\small $\langle \theta \rangle ^\circ$- mean emission angle }
\end{table}
As one can see, even tough the values change for the different parameters, for all cases an increase in the strength of the gravitational field (decrease in $\mu$) leads to an increase in the mean production angle. We note that the two intermediate values may not be very accurate because of the errors in the numerical integration, but they are in conformity with the overall trend.
\subsection{Flat space limit}
As we have mentioned, this process is forbidden in flat space because of incompatibility with energy-momentum conservation. This implies that the probability in dS space must vanish in the limit $\omega \rightarrow 0$. To get a better picture of the behaviour, we keep the leading term in $\mu = m/\omega$ in the probability, and seek the approximate expression around three different angular configurations between the involved momenta. \par
The main contribution to the probability in the flat limit comes from the $g_k$ term in (\ref{PP}). This is not easy to see however, because the $\mu$-dependence is intertwined with the angular dependence. Using the approximate relations
\bq
 \nu \simeq i\mu,\quad \quad \sinh(\pi\mu) \simeq \cosh(\pi\mu) \simeq \frac{1}{2}\,e^{\pi\mu},
 \eq the leading contribution to the probability becomes:
\bq
\label{LT}
\mathcal{P}(\vec{p},\vec{p}\,',\vec{k}) &=& \frac{e^2\pi^2}{16}\frac{p'^2p^2\sin^2(\chi)}{(2\pi)^6k^3}\ \vert\, g_\nu(p',p,k+i\epsilon)\, \vert^{\,2} \\
&=& \frac{e^2\pi^2}{4}\frac{\sin^2(\chi)}{(2\pi)^6kpp'} \ \left\vert\, ie^{-\pi\mu}\, _2F_1\left(\frac{3}{2} -i\mu,\frac{3}{2} +i\mu, 2, \frac{1+\cos(\chi)}{2} +i\zeta(\epsilon)\right) \right. \nn \\
&\times& \left(\mu^2 + \frac{1}{4}\right)^2 \ \  + \ \, \left.\, e^{-2\pi\mu}\,_2F_1\left(\frac{3}{2} -i\mu,\frac{3}{2} + i\mu, 2, \frac{1-\cos(\chi)}{2}-i\zeta(\epsilon)\right)\,\right\vert^{\,2}, \nn
\eq
where we have denoted $\zeta(\epsilon) = \frac{\epsilon k}{2pp'}$. \\
The remaining terms fall off as higher powers of $e^{-\pi\mu}$. We have kept track of the $\epsilon$ parameter because it plays a crucial role in the form of the probability, as we shall see below. \par
Using the formulas from Appendix (\ref{A.3}) to approximate the hypergeometric functions contained in (\ref{LT}), we obtain the expression of the probability in the particular cases $\chi = {0,\pi/2,\pi}$.
\begin{eqnarray}
\mathcal{P}(\chi \simeq 0)\ \ \ &\simeq&\ \ \frac{1}{(2\pi)^6}\,\frac{\,e^2}{kpp'}\,\frac{\,\chi^2}{\chi^4 + 16\zeta(\epsilon)^2} \label{P0}\\
\mathcal{P}(\chi = \pi/2) &\simeq&\ \ \frac{1}{(2\pi)^6}\,\frac{e^2}{kpp'}\,\frac{\pi\mu}{2}\, e^{-\pi\mu} \\
\mathcal{P}(\chi \simeq \pi)\ \ \ &\simeq&\ \ \frac{1}{(2\pi)^6}\, \frac{e^2}{kpp'}\,\frac{(\pi-\chi)^2}{\left(\pi-\chi\right)^4+16\zeta(\epsilon)^{2}}\ e^{-2\pi\mu}
\end{eqnarray}
The difference among the three particular angular configurations is striking. For the one-photon pair production process in dS space we have found that the probability falls of as $e^{-\alpha(\chi)\mu}$, with $\alpha(\chi) >0$. This behaviour carries over to other tree-level processes, and most likely also to higher order processes. Most importantly, $\alpha \rightarrow 0$ for $\chi\simeq\theta \rightarrow 0$, which means that small angle pair production is dominant in the small expansion parameter limit. \par
By extrapolating (\ref{P0}) to a small vicinity around $\theta = 0$, we conclude that because here the $\mu$ dependence is weak, the probability can remain large even for a weak gravitational field. The probability for the process, being a function of the ratio $\mu = m/\omega$, given a small enough mass for the scalar field, can be significant even for the present day expansion. Our results then suggest that this can still be the case even for particles with Compton wavelengths much smaller than the Hubble radius ($m \ll \omega$). This could have potentially interesting astrophysical consequences.
\section{Conclusions}
We have studied the first order QED process of one-photon scalar pair production, on the expanding Poincare patch of the de Sitter spacetime. \par
The transition probability was evaluated with three different methods for three different domains of the expansion parameter: a) by approximating the analytical expression in the weak field case ($ m \gg \omega$), b) by numerical integration for the strong gravitational domain $( m \sim \omega)$ and c) by analytical integration, facilitated by the simple form of the mode functions for a particular strong field case ($ m = \sqrt{2} \omega$).
The coherence of the results acts as check for the validity of the methods. \par
The resulting probability distribution was represented as a surface in the momentum space (fig.\ref{2}). We have found that there is a moderate interaction (i.e. energy exchange) with the background:
\begin{itemize}
\item in general the probability is concentrated around the configuration that is closest to the energy conservation condition ($k \simeq p + p', \chi \simeq \theta \rightarrow 0$)
\item by turning up the strength of the gravitational field, the production at intermediate angles is enhanced, while remaining vanishingly small for backward scattered and high momentum pairs.
\end{itemize} \par
To make the above description more precise, we have calculated the mean emission angle for various configurations and have found accordingly that it increases/decreases with the increase/decrease of the expansion parameter. In the flat limit the mean angle goes to zero and probability vanishes as expected. Surprisingly, the fall of the probability is rather mild for small angle production. \par
To investigate this behaviour, we have approximated the expression of the probability in the vicinity of different angular configurations, while keeping only the leading term in $\mu = m/\omega$. We have found that the the probability falls off as $e^{-\alpha(\chi)\mu}$, with $\alpha(\chi) >0$. This behaviour carries over to other tree-level processes, and probably also to higher order processes. Most importantly, $\alpha \rightarrow 0$ for $\chi\simeq\theta \rightarrow 0$, which means that small angle pair production is dominant in the small expansion parameter limit. Noting that the probability is a function only of the ratio $m/\omega$, given a small enough mass the probability can be large even for the present day expansion. Our results then seem to suggest that even for a scalar field with Compton wavelength much smaller than the Hubble radius ($m\sim\omega$ in natural units), the probability of pair production around small angles can remain non-negligible for the present expansion. The fact that we do not observe this process for example for the CMB photons, gives a lower bound for the possible masses of such ultra-light particles. Such scalar fields, with Compton wavelengths between Hubble and galactic order, have been popular in recent years in beyond the standard model physics, as candidates for dark matter and dark energy. \par
We signal that these small angle results have to be taken with a dose of skepticism. This is because the quantitative description is highly dependent on the switch-off parameter $\epsilon$. When the vanishing limit is taken the probability is divergent. This is not however a dangerous divergence, being rather a universal trait of theories where massless particles are involved. It arises also in flat space, for example, in the case of a radiating charge kept at constant acceleration by an external source or evolving in the field of a nucleus (bremsstrahlung). \par
It would be interesting to see whether this divergence can be dealt with in a similar manner as in the flat space counterparts. Indeed the entire divergence structure of these transition probabilities is interesting and might have significant consequences.
\begin{description}
\item[a] for the bremsstrahlung case, the (IR) divergence is eliminated when the higher order contributions and also the elastic scattering are summed. It would be interesting to see whether the $\epsilon \rightarrow 0$ divergence in the de Sitter amplitudes might also be eliminated when the higher order interactions and also the elastic scattering off the gravitational field are taken into account.
\item[b] the (UV) divergence noted in section \ref{sec 3.3} is more bothersome. It arises because of the finite (conformal) time integration in the amplitude, and we have argued that it is similar to the transient effects arising when one performs a sudden decoupling of the fields in a process on flat space. This might represent a breakdown of the applicability of the \emph{eternal} dS space in physical contexts.
\item[c] the dS QED probabilities have an additional divergence in the $ m/\omega = \{ 0, 1.5 \} $ domain,  stemming from $\eta \rightarrow 0$ end of the temporal integral. This is the famous (IR) divergence of dS space which is relevant for inflation, and is still an open problem to date (for a comprehensive list of references see \cite{A1}). It might be worth investigating if the interaction could have a word to say regarding this issue.
\end{description}

\begin{acknowledgments}
We are grateful to Ion Cotaescu, Cosmin Crucean and Nistor Nicolaevici for useful discussions and constructive criticism that led to the improvement of this work. \\
This work was supported by the strategic grant POSDRU/159/1.5/S/137750, Project ”Doctoral and Postdoctoral programs support for increased competitiveness in Exact Sciences research” cofinanced by European Social Fund within the Sectoral Operational Programme Human Resources Development 2007-2013.
\end{acknowledgments}

\appendix

\section{Mathematical Toolbox}
\label{A}
\subsection{Bessel functions}
\label{A.1}
Here we will list the relations and integrals involving Bessel and Hankel functions, that enter in the calculation of the amplitude. \cite{GR,PB} \\
For the particular value of the index $\pm\frac{1}{2}$ the expression of the Hankel functions is:
\bq \label{H12}
H^{(1)}_{\frac{1}{2}} (z) =  -i\sqrt{\frac{2}{\pi z}}\,e^{\,iz}, &\qquad& H^{(2)}_{\frac{1}{2}}(z) = i\sqrt{\frac{2}{\pi z}}\,e^{-iz} \nn \\
H^{(1)}_{-\frac{1}{2}} (z) =\,\ \ \  \sqrt{\frac{2}{\pi z}}\,e^{\,iz}, &\qquad& H^{(2)}_{-\frac{1}{2}}(z) =\ \sqrt{\frac{2}{\pi z}}\,e^{-iz}
\eq
The Hankel functions can be expressed in terms of Bessel J functions as follows:
\begin{eqnarray}
H^{(1)}_{\nu}(z)&=&\frac{J_{-\nu}(z)-e^{-i\pi\nu}J_{\nu}(z)}{i\sin{\pi\nu}} \\
H^{(2)}_{\nu}(z)&=&\frac{e^{i\pi\nu}J_{\nu}(z)-J_{-\nu}(z)}{i\sin{\pi\nu}} .\label{a3} \nn
\end{eqnarray}
The temporal integral can be thus turned into 4 integrals of the form:
\begin{equation}
\int\limits_{0}^{\infty}d\eta\,\eta\,J_{\pm \nu}(p\,\eta)J_{\pm \nu}(p'\eta)\,e^{ik\eta}\,\,\,\,\,\,\,\,\,\,\, \int\limits_{0}^{\infty}d\eta\,\eta\,J_{\pm \nu}(p\,\eta)J_{\mp \nu}(p'\eta)\,e^{ik\eta}.
\end{equation}
The first type of integral, which contains Bessel functions with equal sign indices, can be solved by using:
\bq
\int\limits^\infty_0 d\eta\ \eta\,J_{\pm\nu} (p\,\eta)\, J_{\pm\nu}(p'\eta)\,e^{ik\eta - \epsilon\eta} &=&- \frac{\,ik\ \,}{\pi(p p')^{3/2}}\,\frac{d}{dz} Q_{\pm\nu - \frac{1}{2}} \left(z\right)
\eq
\bq
\frac{d}{dz}Q_{\nu}(z)&=& \frac{\pi\nu(\nu+1)}{4\sin{\pi\nu}}\left[e^{\mp i\pi\nu}\,_{2}F_{1}\left(1-\nu,2+\nu;2;\frac{1-z}{2}\right) \right.\\
&& \left. \qquad \qquad \quad \ + \quad \,_{2}F_{1}\left(1-\nu,2+\nu;2;\frac{1+z}{2}\right)\right] \nonumber \label{a6},
\eq
where $z = \frac{p^2 + p'^2 - (k+i\epsilon)^2}{2pp'}$, and the $\pm$ branch is selected according to the sign of -Im(z).
The second type of integral, containing Bessel functions with opposite sign indices, is given by:
\bq
\int\limits^\infty_0 d\eta\ \eta\, J_{\,\nu} (p\,\eta)\, J_{-\nu}(p'\eta)\,e^{ik\eta - \epsilon\eta} =
\eq
\bq
= \left(\frac{p}{p'}\right)^\nu\,\left(\frac{1}{ik}\right)^2\, \frac{\sin(\pi\nu)}{\pi\nu\,} \, F4\left(1,\frac{3}{2},1+\nu,1-\nu,\frac{\,p^{2}}{(k+i\epsilon)^2}, \frac{\,p'^2}{(k+i\epsilon)^2}\right)  \nn
\eq
With the above formulas, the temporal integral can be written as:
\begin{eqnarray}
e^{-i\pi\nu}I^{(2,2)}_{\nu}= [g_{\nu}(p,p\,',k)+ g_{-\nu}(p,p\,',k)+h_{\nu}(p,p\,',k)+h_{-\nu}(p,p\,',k)]. \label{E9},
\end{eqnarray}
where we have introduced the notations
\bq
\textmd{g}_{\pm \nu}(p,p\,',k)&=& \frac{ik}{4(pp\,')^{3/2}}\frac{\left(\nu^{2}-\frac{1}{4}\right)e^{\mp i\pi \nu}}{\cosh(\pi \nu)\sinh^{2}(\pi k)}\left[i e^{\mp i\pi \nu}\,_{2}F_{1}\left(\frac{3}{2}\pm \nu,\frac{3}{2}\mp \nu;2;\frac{1-z}{2}\right)\right.\nonumber\\
&&\left. \qquad \qquad + \quad \, _{2}F_{1}\left(\frac{3}{2}\pm \nu,\frac{3}{2}\mp \nu;2;\frac{1+z}{2}\right)\right] \nn \\
\textmd{h}_{\pm \nu}(p,p\,',k) &=& -\frac{k^{-2}}{\pi \nu\sinh(\pi \nu)}\left(\frac{p}{p\,'}\right)^{\pm \nu}F_{4}\left(\frac{3}{2}, 1, 1\pm \nu, 1\mp \nu; \frac{p^{2}}{k^{2}}, \frac{p\,'^{2}}{k^{2}}\right).
\eq

\subsection{$_2F_1$ hypergeometric function}
\label{A.3}
In order to find the asymptotic form of the probability, in the flat limit, we need to find an approximate form for the Gauss hypergeometric functions that are contained in the leading terms. More explicitly we wish to find the form of $_2F_1\left(\frac{3}{2} - \nu, \frac{3}{2}+\nu,2,x\right)$, for the distinct cases $x=\{0,\frac{1}{2},1\}$, in the limit $\mu \rightarrow \infty, \nu \rightarrow i\mu$. \\ \\
With the use of the relations \cite{GR}
\bq
_2F_1(\alpha,\beta,\gamma,x) &=& (1-x)^{\gamma-\alpha-\beta}\, _2F_1(\gamma-\alpha,\gamma-\beta,\gamma,x) \\
_2F_1(\alpha,\beta,\gamma,1) &=& \frac{\Gamma(\gamma)\Gamma(\gamma-\alpha-\beta)}{\Gamma(\gamma-\alpha)\Gamma(\gamma-\beta)},\quad Re(\gamma) > Re(\alpha+\beta),
\eq
we find:
\bq
\label{A31}
_2F_1\left(\frac{3}{2} - \nu, \frac{3}{2} + \nu, 2,x \simeq 1\right) &\simeq& \left(1-x\right)^{-1}\,_2F_1\left(\frac{1}{2}+i\mu, \frac{1}{2} - i\mu, 2,1\right) \\
&=& \left(1-x\right)^{-1} \frac{\Gamma(2)\ \Gamma(1)}{\Gamma\left(\frac{3}{2} + i\mu\right)\Gamma\left(\frac{3}{2} - i\mu\right)} \nn\\
&=& \frac{(1-x)^{-1}}{\left(\frac{1}{2} + i\mu\right)\Gamma\left(\frac{1}{2} + i\mu\right)\left(\frac{1}{2} - i\mu\right)\Gamma\left(\frac{1}{2} -i\mu\right)} \nn\\
&=& \frac{(1-x)^{-1}}{\left(\frac{1}{4}+\mu^2\right)\ \left\vert\,\Gamma\left(\frac{1}{2} + i\mu\right) \right\vert^{\,2}} \nn\\
&=& \frac{1}{(1-x)}\ \frac{\cosh(\pi\mu)}{\pi\left(\frac{1}{4}+\mu^2\right)} \nn
\eq \\
For the case $x = \frac{1}{2}$, we require the formula:
\bq
\label{AB}
_2F_1\left(2\alpha,2\beta, \alpha+\beta + \frac{1}{2}; \frac{1-\sqrt{y}}{2}\right) = A\, _2F_1\left(\alpha,\beta, \frac{1}{2};y\right) + B\,\sqrt{y} \,_2F_1\left(\alpha+\frac{1}{2}, \beta + \frac{1}{2}, \frac{3}{2}; y\right) \nn
\eq
\bq
A=\frac{\Gamma\left(\alpha+\beta+\frac{1}{2}\right)\ \sqrt{\pi}}{\Gamma\left(\alpha+\frac{1}{2}\right) \Gamma\left( \beta + \frac{1}{2}\right)},\qquad
\quad \alpha = \beta^* = \frac{3}{4} + \frac{i\mu}{2},
\eq
In our case y = 0, and the right-hand side of (\ref{AB}) reduced to A. \\
Using also the limit:
\bq
\lim_{\vert y\vert \rightarrow \infty}\ \vert \Gamma(\alpha + i\beta) \vert\ e^{\frac{\pi \vert \beta\vert}{2}}\ \beta^{\left(\frac{1}{2} - \alpha\right)} = \sqrt{2\pi},
\eq
we obtain:
\bq
\label{A32}
_2F_1\left(\frac{3}{2} - \nu, \frac{3}{2} + \nu, 2, x = \frac{1}{2}\right) &=& \frac{\sqrt{\pi}}{\Gamma\left(\frac{5}{4} +\frac{i\mu}{2}\right)\Gamma\left(\frac{5}{4} - \frac{i\mu}{2}\right)} \\
&=& \frac{\sqrt{\pi}}{\left(\frac{1}{4} - \frac{i\mu}{2}\right)\Gamma\left(\frac{1}{4} + \frac{i\mu}{2}\right)\left(\frac{1}{4} + \frac{i\mu}{2}\right)\Gamma\left(\frac{1}{4} - \frac{i\mu}{2}\right)} \nn \\
&\simeq& \frac{\sqrt{\pi}}{\frac{1}{4}\left(\frac{1}{4} + \mu^2\right)2\pi\left(\frac{\mu}{2}\right)^{-1/2}\,e^{-\frac{\pi\mu}{2}}} \nn \\
&=& \sqrt{\frac{4}{\pi}} \left(\mu^2 + \frac{1}{4}\right)^{-1} \left(\frac{\mu}{2}\right)^{1/2} e^{\frac{\pi\mu}{2}} \nn
\eq
Finally:
\bq
\label{A30}
_2F_1\left(\frac{3}{2} - \nu, \frac{3}{2} + \nu, 2, 0\right) =1
\eq

\subsection{Appell $F_4$ function}
\label{A.2}
From the definition of the Appell $F_4$ hypergeometric function, we have:
\bq
F_4(1,3/2,1+\nu,1-\nu,x,y) = \sum_{m,n=0}^\infty \frac{(1)_{m+n}(3/2)_{m+n}}{(1+\nu)_m(1-\nu)_n}\frac{x^m}{m!}\frac{y^n}{n!},
\eq
where the Pochhammer symbol means $(a)_m = a(a+1)...(a+m-1) = \Gamma(a+m)/\Gamma(a)$. \par
We are interested in the limit $\mu \rightarrow \infty, \nu \rightarrow i\mu$. As the parameter $\mu$ appears only in the denominator, it is reasonable to consider the approximation $F_4 \simeq 1$. This is by no means obvious, because our function does not satisfy the absolute convergence criterion: $\sqrt{x} +\sqrt{y} < 1$. \par
\begin{figure}[h]
\centering
\begin{subfigure}{0.5\linewidth}
\centering
\includegraphics[width=3in]{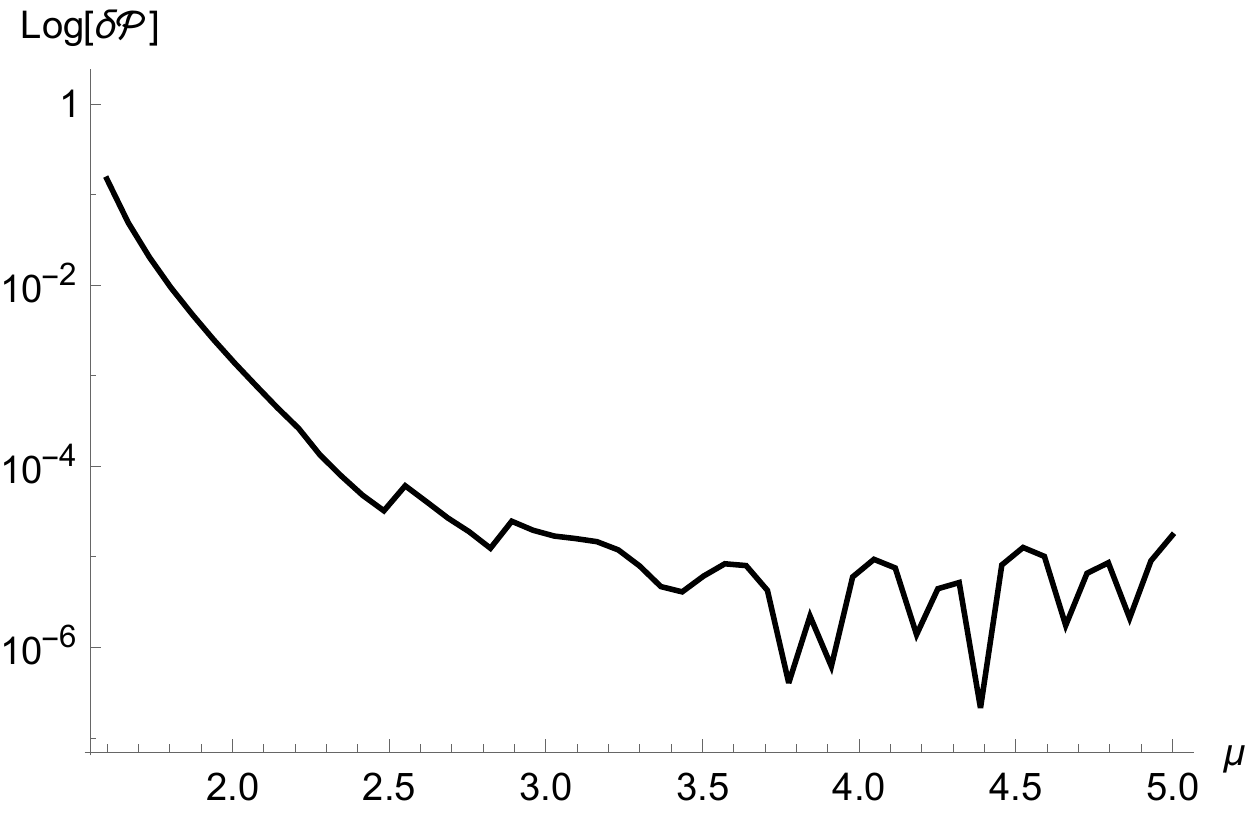}
\caption{}
\label{10a}
\end{subfigure}%
\begin{subfigure}{0.5\linewidth}
\includegraphics[width=3in]{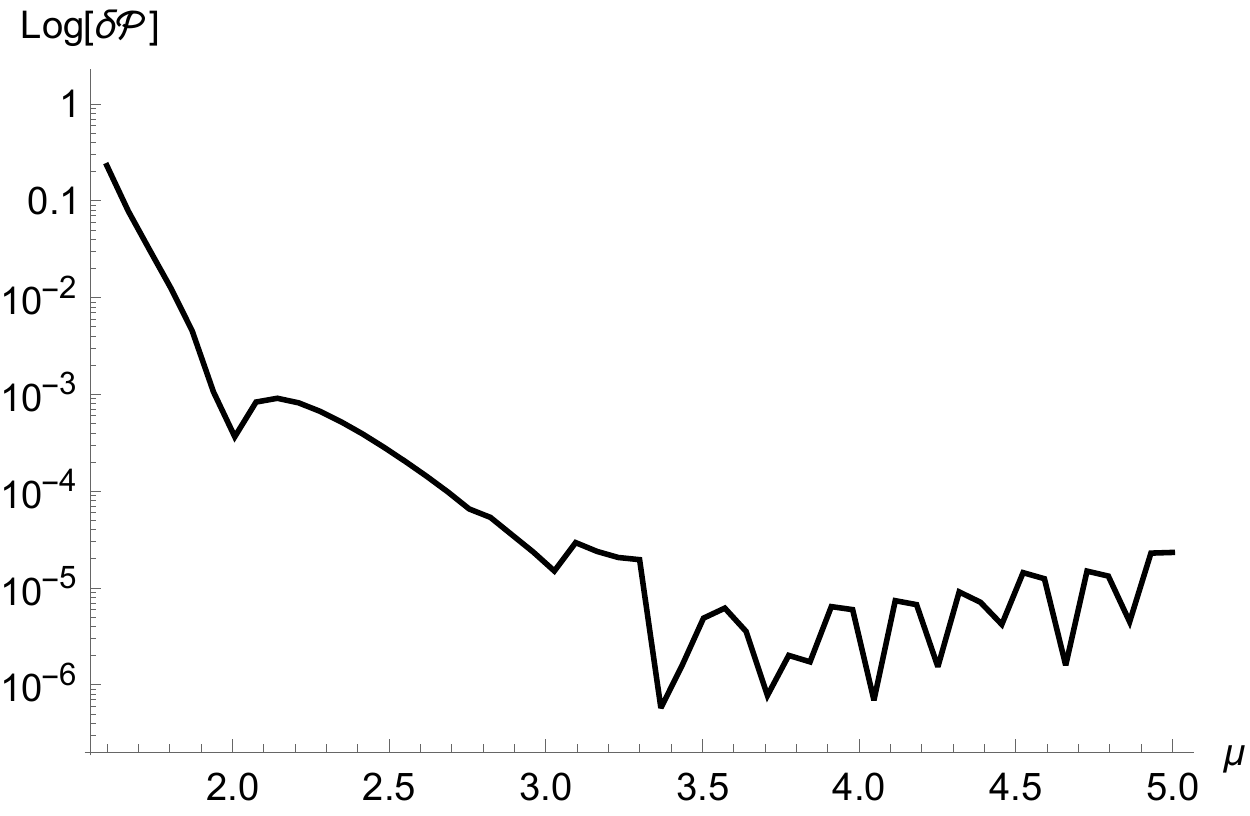}
\label{10b}
\caption{}
\end{subfigure}
\caption{\small The relative error ($\delta\mathcal{P}$) of the approximated to the numerically evaluated the probability, for $k=1,\epsilon= 10^{-2}$ with (a) $p=0.1$  and (b) $p=0.01$ .}
\label{10}
\end{figure}

A comparison of the approximated against the numerically evaluated probability can be seen in fig \ref{10}. We have found that for $m > 5\omega$ this approximation is very good. However, increasing the switch-off parameter $\epsilon$ decreases the accuracy of the approximation.

\section{Added-up probability}
\label{C}
The method of added-up probabilities, introduced by Audretsch and Spangehl in ref.\cite{AS1}, gives a nice way of interpreting in-in amplitudes in terms of quantities, the definition of which is invariant to the nature of the out state. The problem to which it offers a solution is the following: in a setting with interacting fields on a dynamical background, when calculating scattering amplitudes in the S-matrix approach, the effects of the interaction can not, in general, be separated from the cosmological pair creation. \par
In a quasi-flat spacetime, where the cosmological particle production is negligible, one can do the usual Minkowskian perturbation theory. On the other hand, for strong gravitational fields it is not clear whether and how one can neglect or subtract the effects of the cosmological particle production. An alternative possibility is to work with a notion of global measuring apparatus as in ref.\cite{AS4}, and interpret the in-in probabilities as the sole contribution of the mutual interaction, separate from the pure cosmological pair production. \par In the particular case when one of the fields is conformal, for example the Maxwell field in QED, the definition of the vacuum is unambiguous. This means that quanta of this field cannot be created freely from the vacuum, which in turn means that they are good indicators for the effects of the interaction, as the authors of ref.\cite{AS1} point out. The physically measurable quantity then, they argue, is the probability of measuring a certain final state for the conformal field, regardless of the state of the other fields, given a fixed initial configuration. In QFT language this means that the summed probability is obtained by summing and integrating over the complete Fock space of the other fields in the final state. If we want for example, in the context of tree-level scalar QED, to calculate the summed probability for finding no photons, given a photon in the initial state, the summed probability is:
\bq
\label{C.1}
w^{\,add}_{\,\gamma\, \rightarrow\,\#}(p,p',k) &=& \sum_\#\ \left\vert\, \langle\, 0_M;\,\#\, \vert\,S^{(1)}\vert\,  1_{(\vec{k},\lambda)};\,0_\varphi\,\rangle\, \right\vert^{\,2} \\
&=&  \sum_\#\ \langle\, 1_{(\vec{k},\lambda)};\, 0_\varphi\, \vert\, S^{\dag (1)}\,\vert\, 0_M;\,\#\rangle\,\langle\, 0_M;\,\# \, \vert\,S^{(1)}\,\vert\, 1_{(\vec{k},\lambda)};\,0_\varphi\,\rangle, \nn
\eq
where $0_M$ and $0_\varphi$ represent the vacuum states of the Maxwell and scalar field, and \# stands for the complete out Fock space for the scalar field.
\bq
\sum_\# = \int d^3p_1 + \iint d^3p_1 d^3p_2 + ... = \sum_n \idotsint \prod_{i=0}^n d^3p_i
\eq
We observe by writing $\vert 0_M, \# \rangle = \vert 0_M \rangle \bigotimes \vert \# \rangle$ in (\ref{C.1}), that we have obtained an identity operator for the scalar sector. We can thus insert any orthonormal base in place of $\#$, in particular we can insert an in base.
\bq
\sum_\#  \vert \# \rangle \langle \# \vert\quad   \rightarrow \quad  \sum_{in}  \vert in \rangle \langle in \vert
\eq
This turns out to be the best solution for calculating such probabilities, as the summed probability turns into a finite sum of in-in probabilities. In the particular case of (\ref{C.1}), the only term that contributes to the sum is the pair production process (fig. \ref{1}).
\
\bq
w^{\,add}_{\,\gamma\, \rightarrow\,\#} = \left\vert\, \langle\, 1_{(\vec{p}\,)},\tilde{1}_{(\vec{p}\,')}\, \vert\,S^{(1)}\vert\,  1_{(\vec{k},\lambda)}\,\rangle\, \right\vert^{\,2}
\eq
This means that the in-in probability  for one-photon pair production is equivalent, through the notion of the summed probability, with the probability that a photon is absorbed, regardless of the mechanism (regardless of "into what" it is absorbed). \par
It may seem that this is not saying much, but we must note that the second part of the above statement describes a quantity which makes no assumption on the nature of the out state of the scalar field. \par
As a further example, we evaluate the summed probability that is correlate to the one-photon annihilation process, the time-inversed version of fig.(\ref{1}). This is interpreted as the probability of measuring a photon with momentum $\vec{k}$ in the out state (irrespective of the state of the scalar field), given a pair of scalar particles in the initial state.
\bq
w^{\,add}_{\,\varphi\, +\, \varphi^\dag \,\rightarrow\,\gamma\, +\,\#} &=& \sum_\#\ \left\vert\, \langle\, 1_{(\vec{k},\lambda)};\,\#\, \vert\,S^{(1)}\vert\,  0_M;\, 1_{(\vec{p}\,)},\tilde{1}_{(\vec{p}\,')}\,\rangle\, \right\vert^{\,2} \\
&=& \left\vert\, \langle\, 1_{(\vec{k},\lambda)}\, \vert\,S^{(1)}\vert\,1_{(\vec{p}\,)},\tilde{1}_{(\vec{p}\,')}\,\rangle\, \right\vert^{\,2} \nn \\
&+& \left\vert\, \langle\, 1_{(\vec{k},\lambda)};\,1_{({\vec{p}-\vec{k}}\,)},\tilde{1}_{(\vec{p}\,')}\, \vert\,S^{(1)}\vert\,1_{(\vec{p}\,)},\tilde{1}_{(\vec{p}\,')}\,\rangle\, \right\vert^{\,2} \nn \\
&+& \left\vert\, \langle\, 1_{(\vec{k},\lambda)};\,1_{(\vec{p}\,)},\tilde{1}_{(\vec{p}\,'-\vec{k})}\, \vert\,S^{(1)}\vert\,1_{(\vec{p}\,)},\tilde{1}_{(\vec{p}\,')}\,\rangle\, \right\vert^{\,2} \nn \\
&+& \int d^3p''\,\left\vert\, \langle\, 1_{(\vec{k},\lambda)};\,1_{(\vec{p}\,'')},\tilde{1}_{(-\vec{p}\,''-\vec{k})},1_{(\vec{p}\,)},\tilde{1}_{(\vec{p}\,')}\, \vert\,S^{(1)}\vert\,1_{(\vec{p}\,)},\tilde{1}_{(\vec{p}\,')}\,\rangle\, \right\vert^{\,2} \nn
\eq
The first term represents the one-photon annihilation process. The others are emission and triplet production processes, respectively, where the other particles pass through to the final state unchanged.
These are the processes that are indistinguishable from the point of view of a photon counter that measures the number of photons in the final state \cite{AS1}.

\end{document}